\documentclass[aps,prl]{revtex4}
\usepackage{amsmath,epsfig,capt-of,ifthen,calc}

\newcommand{\mref}[1]{(\ref{#1})}
\begin{document}
{\small

\hfill UB-ECM-PF-06/09

\hfill March 2006
}

\title{Monopole-Antimonopole Correlation Functions in 4D $U(1)$ Gauge
Theory}

\author{Luca Tagliacozzo}
\email{luca@ecm.ub.es}
\affiliation{Departament d' Estructura i Constituents de la Mat\`eria,Universitat de Barcelona.\\
647, Diagonal, 08028 Barcelona, Spain.}

\begin{abstract}

 We study the two-point correlator of a modified Confined-Coulomb transition order parameter  in four dimensional compact $U ( 1 )$ lattice gauge theory with Wilson action. Its long distance behavior in the confined phase turns out to be governed by a single particle decay. The mass of this particle is  computed and found to be in agreement with previous calculations of the $0^{++}$ gaugeball mass.
Remarkably, our order parameter allows to  extract a good signal to noise ratio for masses with  low statistics. The results we present provide  a numerical check of a theorem about the structure of the Hilbert space describing the confined phase of four dimensional compact $U (1)$ lattice gauge theories.  
\end{abstract}

 \maketitle

\section{Introduction}
 Quantum Chromo-Dynamics (QCD), the present theory for strong interaction, is formulated in terms of quark and gluon fields. The latter are carriers of the $SU(3)$ symmetric colour interaction.
At low energies, however, the only existing particles are hadrons. To explain this fact, one has to suppose that quarks are confined inside hadrons. The attempt to understand  this  obscure property of quarks, confinement,  has led to study  simplified models, defined on a lattice, but still exhibiting confinement \cite{Wilson:1974sk} . The simplest among such models are Ising systems on  two and three dimensional lattices. They possess global discrete symmetries. Their generalisation to local symmetries are known as gauge Ising models and still exhibit confinement \cite{Balian:1974ts,Balian:1974ir}.
A further step to approach QCD is to consider models with  continuous gauge symmetry in four dimensions (4D).   

Among them is compact $U ( 1 )$ lattice gauge theory (U(1)lgt) \cite{Creutz:1979zg}. This model has two phases:  a confined phase at strong coupling and a Coulomb phase at weak coupling.
The order  and the location of the phase transition  that separate them depend on the  action form and has been a long standing subject of  debate \cite{Lautrup:1980xr,Evertz:1984pn,Arnold:2000hf,Vettorazzo:2003fg}.
 The idea that  monopoles could play a crucial role in the description of confinement appeared after the inspiring work of Polyakov \cite{Polyakov:1975rs} in three dimensions and the contemporary  conjecture, known as dual superconductivity (DS),  by Mandelstam and 't Hooft about confinement in QCD \cite{Mandelstam:1974vf,'tHooft:1975pr}. 
This conjecture states that confinement  could be described by a mechanism  similar to that responsible of  superconductivity.
In the confined phase, the vacuum should be filled with monopoles     in the same way a  superconductor ground state is filled with Cooper pairs.
In the case of U(1)lgt  a finite density of monopoles was  discovered in the confined phase and vanishing  in Coulomb phase \cite{DeGrand:1980eq}.

The introduction of a well defined  order parameter for the Confined - Coulomb transition in  U(1)lgt
\cite{Frohlich:1987er,Polley:1990tf,DelDebbio:1994sx,DiGiacomo:1997sm}  proved, eventually, the condensation  of monopoles at the transition.
  Using the Wilson action, this transition can be shown to be of first order (with very long correlation length) and  located, in the thermodynamic limit,  at $\beta_c \approx 1.011$ (for recent high
precision measurements of the transition point see {\cite{Arnold:2000hf}}).
 It is this  order parameter, a monopole creation operator,   that allows to  rigorously distinguish between the Confined and the Coulomb phases of U(1)lgt. Other equivalent order parameters  have been constructed later \cite{Vettorazzo:2003fg}.

 In  {\cite{DiGiacomo:1997sm}},  a preliminary study of  the mass of the monopole had allowed to claim  that the DS exhibited in the Confined phase of U(1)lgt was of type II. 
 In  {\cite{Jersak:1999nv}}, a similar  study  has been attempted  with the Villain action  but the authors could not gain enough statistics to study the monopole  mass  in the confined phase.

After the work of Seiberg and Witten \cite{Seiberg:1994rs} supporting the DS picture of confinement in the  context of supersymmetric $N=1$ gauge theories,  we  proposed an effective field theory (EFT)  describing  low energy physics for 4D U(1)lgt  in terms of monopole fields
{\cite{Espriu:2003sa}}. The EFT should work regardless of the  action  chosen on the lattice provided  the correlation length is sufficiently large.

In  a later  work, we tried to compare the  predictions of the spectrum extracted from the EFT with the ones available from lattice simulations. This revealed  some inconsistencies in lattice spectra 
{\cite{Espriu:2004ib}}. Relying on a theorem in  {\cite{Frohlich:1986sz}} we tried to match the results of  {\cite{DiGiacomo:1997sm}} with the ones
of  {\cite{Majumdar:2003xm, Cox:1997nq}} (in the latter case  the comparison can only  be qualitative as these results are obtained with Villain action).
This theorem  states that, in the confined phase of 4D U(1)lgt, the Hilbert space of magnetically charged
particles is contained into the Hilbert space of the neutral ones. 
 Masses extracted  from  operators  creating states with the same  quantum numbers (angular momentum, parity and charge conjugation ($J^{PC}$))  but with different magnetic charges should coincide. The monopole mass of {\cite{DiGiacomo:1997sm}} should coincide with the 
$0^{++}$ gaugeball mass of  {\cite{Majumdar:2003xm}} and the dual photon mass of  {\cite{DiGiacomo:1997sm}} should coincide with the $1^{-+}$ gaugeball mass of  {\cite{Majumdar:2003xm}}. 

This, as we already pointed out {\cite{Espriu:2004ib}}, was clearly not the case even at a qualitative level.   
In  {\cite{DiGiacomo:1997sm}}, the monopole was heavier than
the dual photon whereas in  {\cite{Majumdar:2003xm,Cox:1997nq}}
the scalar gaugeball was lighter than the axial vector gaugeball.  

Later on, new results were obtained for the dual photon  mass studying the electric flux tube profile
between static charges {\cite{Panero:2005iu, Koma:2003gi}}. The  results presented therein  (assuming the dual superconductor picture) 
reconcile the dual photon mass with the one of the axial vector gaugeball. The only piece of the spectrum in disagreement with the theorem  was the monopole mass as calculated in  {\cite{DiGiacomo:1997sm}} . 

Our main result is to  reconcile the value of the mass extracted from monopole correlators with the one of the $0^{+ +}$ gaugeball.

For that purpose,  we introduce a modified order parameter as explained  in the  next section. 
Then, we present the detailed analysis of  data obtained simulating this new order parameter on the lattice.
We check that a single particle is responsible for the decay of its two-point correlation function. 
We  extract  the  mass of this particle and find complete agreement with that of the $0^{+ +}$ gaugeball. 
Our method allows to extract a very good signal to noise ratio for the masses with an amount of data one order of magnitude smaller than that needed for existing techniques.
We conclude with a discussion of the physical scenario emerging from our study.
\section{The operator for spectral studies}
The partition function of 4D U(1)lgt  is defined as:
\[ Z ( \beta ) = \int \left( \prod_{( \vec{n}, t )} \prod_{\mu = 0} ^ 3 \mathcal{D}
   \theta_{\mu}(\vec{n},t) \right) \exp\left( - \beta  S \right). \]
We chose to consider the  Wilson action:
\begin{equation} 
S = -\sum_{(\vec{n}, t)}\sum_{i>\mu = 0} ^3 \left( \cos d \theta_{i\mu}(\vec{n},t) - 1\right). \label{wilson}
\end{equation} 
where $\theta_{ \mu}(\vec{n},t)$ are the link variables of the
four dimensional lattice whose sites are labelled by $(\vec{n},t)$ . We use  $i$  as index for spatial directions and  $\mu$ as index that spans all the four directions of the lattice. The field $\theta_{\mu}(\vec{n},t)$ take value in $U ( 1 )$ and $\prod_{( \vec{n}, t )}\prod_{\mu = 0 }^ 3\mathcal{D}
   \theta_{ \mu }(\vec{n},t)$ is the Lesbegue measure for each variable.  We will abbreviate it  as $\left(\prod \mathcal{D} \theta \right)$.
$d \theta_{i\mu}(\vec{n},t)$ is the lattice field strength  term obtained acting with the exterior derivative on the link variables. $d \theta_{i\mu}(\vec{n},t)$ is hence defined  on plaquettes identified by the coordinates $(\vec{n},t)$  lying on the plane $i-\mu$.
 At last, $\beta=\frac{1}{g^2}$ with  $g$  being the $U(1)$ the coupling constant.

The order parameter  for the Coulomb-Confined transition we consider was introduced by the Pisa group 
{\cite{DelDebbio:1994sx,DiGiacomo:1997sm}}. It is the mean value of a monopole creation
operator and shifts the plaquette field strength at a given Euclidean time by the contribution of the vector
potential produced by a static magnetic source $\vec{b}( \vec{x} )$.
In the continuum, using standard notation this would be:
\begin{equation}
 \mu (\vec y, t) |\vec A(\vec x,t) \rangle\equiv
 |\vec A(\vec x,t) + \frac{1}{e}\vec b(\vec x-\vec y)\rangle\end{equation}
with
\begin{equation}
\mu (\vec y,t) = \exp\left[i \frac{1}{e}\int {\rm d}^3 x\,\vec E(\vec x,t)
\vec b(\vec x-\vec y)\right]. 
\end{equation}
Further details on the subject can be found  directly
in  {\cite{DiGiacomo:1997sm}}. 

The lattice version of the operator is defined as:
\begin{equation}
\mu (\vec y,t)=  \exp \sum_{\vec{n}}\sum_{ i = 1} ^ 3 \beta \left( \cos
  ( b_i ( \vec{y}-\vec{n}) - d \theta_{i 0}( \vec{n},t ) ) -\cos(d \theta_{i 0}( \vec{n},t )) \right)\label{mulat}
\end{equation}
If we  consider the effect of a monopole antimonopole pair placed at points $(\vec{x},t_1)$ and $(\vec{y},t_2)$ on the lattice  the corresponding correlator will be :
\begin{eqnarray}
  \langle \mu ( \vec{x}, t_1 ) \bar{\mu} ( \vec{y}, t_2 ) \rangle & = &
  \frac{1}{Z} \int \left(\prod \mathcal{D} \theta \right) \prod_{( \vec{n}, t )}\prod_{ i>\mu = 1} ^ 3 \exp  \beta \left(
  \cos d \theta_{i\mu}( \vec{n}, t ) - 1 \right) \label{mu} \nonumber\\
  &  & \prod_{\vec{n}, t \notin \{ t_1, t_2 \}}\prod_{ i = 1} ^ 3 \exp  \beta \left(
  \cos d \theta_{ i0}( \vec{n}, t ) - 1 \right) \\
  &  & \prod_{\vec{n}, t \in \{ t_1, t_2 \}}\prod_{ i = 1} ^ 3  \exp  \beta \left( \cos
  ( b_i ( \vec x, \vec y ,\vec{n}) - d \theta_{i 0}( \vec{n},t ) ) - 1 \right). \nonumber
\end{eqnarray}
In this formula $b_i ( \vec{x}, \vec y ,\vec{n})=b_i ( \vec{x}-\vec{n})_{t=t_1}-b_i ( \vec{y}-\vec{n})_{t=t_2}$ is  the $i$ component of the lattice  vector potential produced by  the  monopole anti-monopole pair. It has  support
only on the two time slices $t_1$ and $t_2$.  
 From now on we will consider the case for which $\vec{x} =
\vec{y}$ and the external field will be denoted $b_i ( \vec{x}-\vec{n})$.

The expectation value \mref{mulat} is very difficult to extract from Monte-Carlo simulations. In order to see why this is so we  define a new operator

\begin{equation} \mathcal{O}( \Delta t ) = \prod_{\vec{n} ,t \in\{ t_1,t_2\}}\prod_{ i = 1} ^ 3  \exp
    \beta \left( \cos ( b_i (\vec x- \vec{n} ) - d \theta_{i 0}( \vec{n},t ) ) - 1
   \right)\label{O}
\end{equation}
 where, to simplify the notation, we have ignored its dependency on all variables but $\Delta t=|t_2-t_1|$ . This is  justified by the fact that we will always deal with expectation values of \mref{O} that really  depends (among other variables) on  $\Delta t$. We   introduce  a modified action:
\begin{equation}
 \tilde{S} = - \sum_{(\vec{n}, t)}\sum_{ i>\mu = 1}^ 3 \left( \cos d \theta_{i
   \mu} (\vec{n}, t)- 1 \right)-\sum_{\vec{n}, t \notin \{ t_1, t_2 \}}\sum_{ i = 1}^ 3  \left(
   \cos d \theta_{i 0}(\vec{n}, t) - 1 \right) .\label{tilds} 
\end{equation} 
It is important to notice that the action \mref{tilds} differs from the
standard action \mref{wilson} only on the two time-slices $t_1$ and $t_2$ where the monopole
operator has support. On these time-slices \mref{tilds} vanishes.

We can express the two point correlation function \mref{mu} as the mean value of the
operator \mref{O} on configurations generated with \mref{tilds}:
\begin{equation}
 \langle \mu ( \vec{x}, t_1 ) \bar{\mu} ( \vec{x}, t_2 ) \rangle =\frac{1}{Z} \int \left(\prod \mathcal{D} \theta \right) \exp \left( - \beta \tilde{S} \right) \mathcal{O}( \Delta t )
   \label{o}
\end{equation}
In this way the difficulty in measuring \mref{mu} appears immediately: 
we are trying to extract the mean value of \mref{O} on  configurations sampled uniformly
 on its support. On the time-slice $t_1$ and $t_2$, support of \mref{O}, the term $ -\cos(d \theta_{i 0}( \vec{n},t ))$ coming from  \mref{mulat}  cancels the analog  contribution coming from  the action \mref{wilson}. 
In this way, on these time slices,  importance sampling   is  lost. 

This is one of the reasons that forced the
authors of  {\cite{DiGiacomo:1997sm}} to  introduce:
\begin{equation}
  \rho = - \frac{\partial}{\partial \beta} \log \langle \mu ( \vec{x},
  t_1 ) \bar{\mu} ( \vec{y}, t_2 ) \rangle \label{ro}
\end{equation}
With this definition one recovers the importance sampling on the time-slices
$t_1$ and $t_2$. On them the action becomes:
\begin{equation}
 S' ( t_1, t_2 ) = \sum_{\vec{n}, t \in\{t_1, t_2\} }\sum_{ i=1} ^3  \cos ( b_i (\vec x - \vec{n} ) - d \theta_{i 0}(\vec{n}, t) ) - 1 
\label{actprim}
\end{equation}
If one is interested in studying   the mean value of $\rho$ this is enough
{\cite{DiGiacomo:1997sm}}.
In our case we need to go a little further as we want to extract masses in the confined phase.
To do this we are forced to study the decay of \mref{mu} as function of the time separation $t$ between monopole and antimonople.
 Assuming  that for  large $t$ the decay of \mref{mu} is driven by a single particle, in the confined phase, we obtain:
\begin{equation}
  \langle \mu ( \vec{x}, t ) \bar{\mu} ( \vec{x}, 0 ) \rangle \sim \mu^2 +
  A M^{1/2}t^{-{3}/{2}}e^{- M t} . \label{dec}
\end{equation}
In this formula $\mu$ is the v.e.v of the monopole operator $\mu (
\vec{x}, t )$. It  is different from zero in the confined phase.
$A$ is the projection ofthe monopole state on the vacuum . $M$ is
the monopole mass. In order to keep formulas compact  we are neglecting the effect of  periodic
boundary conditions but we will add them when studying numerical results. 

Fluctuations of \mref{ro}, in the confined phase,  are caused by fluctuation of the v.e.v. $\mu^2$ and  completely screen the decay \mref{dec} we need to unmask to extract $M$. To deal with this in  \cite{DiGiacomo:1997sm} a very huge amount of data was collected.
We can, however, introduce a new definition of  $\rho$ that, following the ideas introduced in 
{\cite{Majumdar:2003xm}} in the context of gaugeball spectroscopy, approximates  the connected part of \mref{mu} and eliminates  $\mu^2$. The new  operator is defined as:
\begin{equation}
  \rho' = \frac{\partial}{\partial t} \log \langle \mu ( \vec{x}, t )
  \bar{\mu} ( \vec{x}, 0 ) \rangle . \label{ron}
\end{equation}
With this definition  importance sampling on
the monopole anti-monopole time slices is kept (as in the case of  \mref{ro}) and fluctuations at $\mu^2$ scale are eliminated. Starting  from \mref{dec}
 we obtain the large $t$ behaviour of  \mref{ron}:
\begin{equation}
  \rho' \sim A M^{1/2} {t^{-{3}/{2}}}{e^{-
  Mt}}\left( - M - \frac{3}{2 t} \right) \left(\mu^2 + A M^{1/2} {t^{-{3}/{2}}}{e^{- Mt}}\right)^{-1} 
  \label{derlog}
\end{equation}
This expression has 3 free parameters: $\mu, M, A$. On a large enough lattice  we can expand
\mref{derlog} in   a regime where
$\mu^2  \gg A M^{1/2} {t^{-3/2}}{e^{- Mt}}$. At first order in $ \frac{A M^{1/2}}{\mu^2 }t^{-{3}/{2}}e^{- Mt}
$ we get from \mref{derlog}:
\[ \rho' \sim \left( - M - \frac{3}{2 t} \right) \frac{AM^{1/2}}{\mu^2 }
   {t^{-{3}/{2}}}{e^{- Mt}}  \]
from which it is easier to extract a precise determination of the mass. However the results presented in this paper are obtained
with the full expression \mref{derlog}.

The simulation algorithm for studying 
$\rho'$  decays is very similar to the one needed to study $\rho$ decays.
Starting from the expression \mref{o} and taking the logarithm we get
\begin{equation}
  \partial_{t } \log < \mu ( \vec{x}, t ) \bar{\mu} ( \vec{x},
  0 ) > = \frac{ \int \left(\prod \mathcal{D} \theta \right)  \exp \left( - \beta \tilde{S} \right)
  \partial_{t} \mathcal{O}( t ) }{\int \left(\prod \mathcal{D} \theta \right) \exp \left( - \beta \tilde{S}
  \right) \mathcal{O}( t_0 )} \label{rop}
\end{equation}
We need to  use the first order approximation in $\partial_t f$: 
\begin{equation} \partial_t e^{f ( t )} |_{t
= t_0} = e^{f ( t )} ( \partial_t f ) |_{t = t_0}
\label{approx}
\end{equation}
 In this way  we can express the derivative as : 
\[
\partial_{t} \mathcal{O}( t) =\mathcal{O}( t ) 
\sum_{\vec{n}}\sum_{i=1}^ 3 \partial_{t} \left( \beta ( \cos ( b_i ( \vec{n} ) - d
\theta_{i 0}( \vec{n}, t_0) ) - 1 \right) 
\]
and we can include the factor  $\mathcal{O}( t )$ in the measure. 
This
defines a new action: \[S'=\tilde{S} + \sum_{\vec{n} }\sum_{i=1}^ 3     \cos ( b_i (
\vec{n} ) - d \theta_{ i0}( \vec{n}, t  ) ) - 1. \] The expression \mref{rop} in
this way becomes:
\begin{equation}
  \partial_{t } \log \langle \mu ( \vec{x}, t) \bar{\mu} ( \vec{x},
  0 ) \rangle = \frac{1}{Z'} \int \left(\prod \mathcal{D} \theta \right)  e^{- \beta S'} \partial_{t} \left( \log
  \mathcal{O}( t )-\beta S(t) \right). \label{numrop}
\end{equation}
 where $S(t)$ is the Wilson action defined in \mref{wilson} restricted on the time slices $t$ and 0.
We hence need   to compute  the mean value of \[ \partial_{t} \left( \log
\mathcal{O}( t )-\beta  S(t)\right)\] on configurations generated with $S'$:
\begin{equation}
\rho'=\langle \partial_{t} \left( \log
\mathcal{O}( t )-\beta  S(t)\right) \rangle'\label{numexpr}
\end{equation} 

\begin{figure}\begin{center}\resizebox{!}{5cm}{\epsfig{file=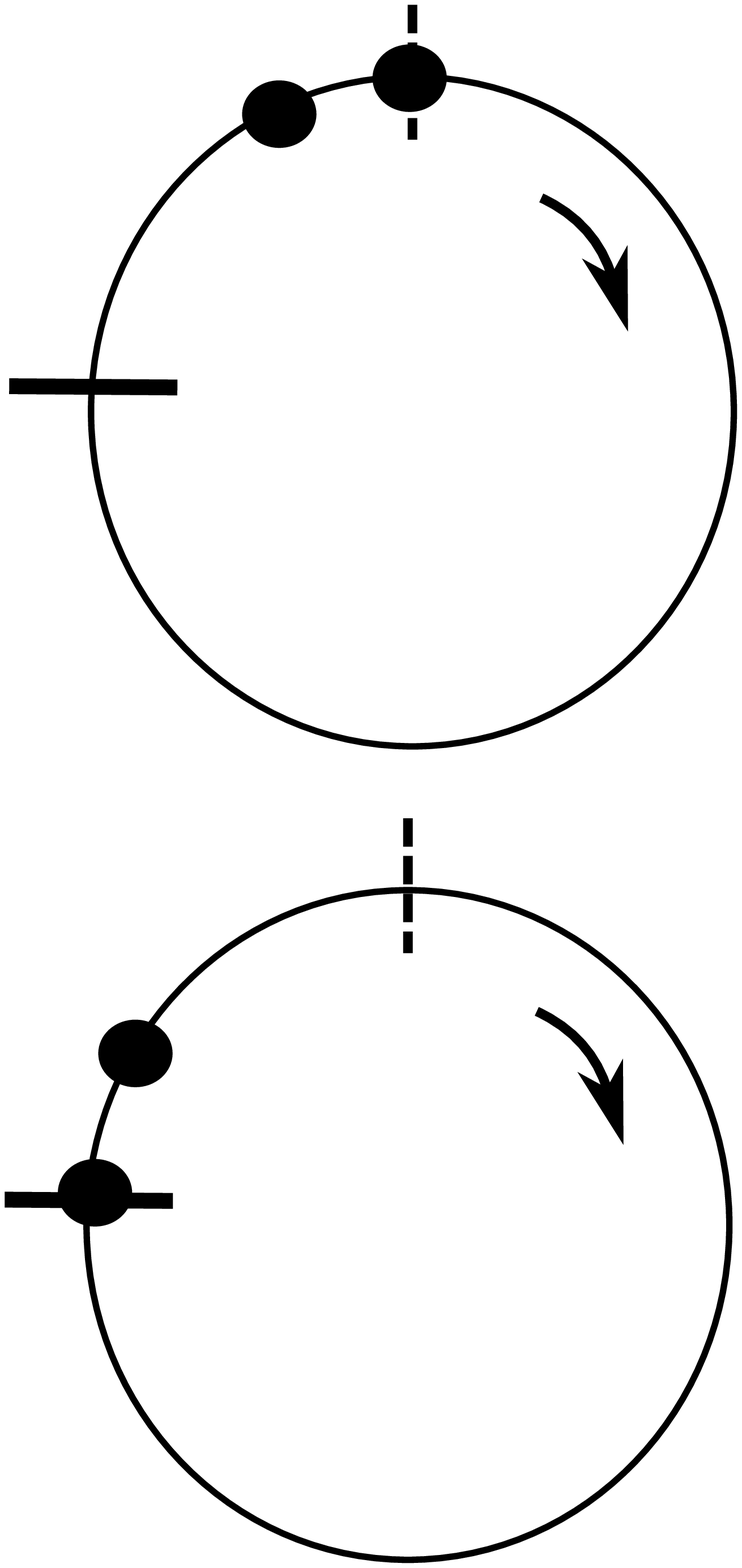}}
\resizebox{!}{5cm}{\epsfig{file=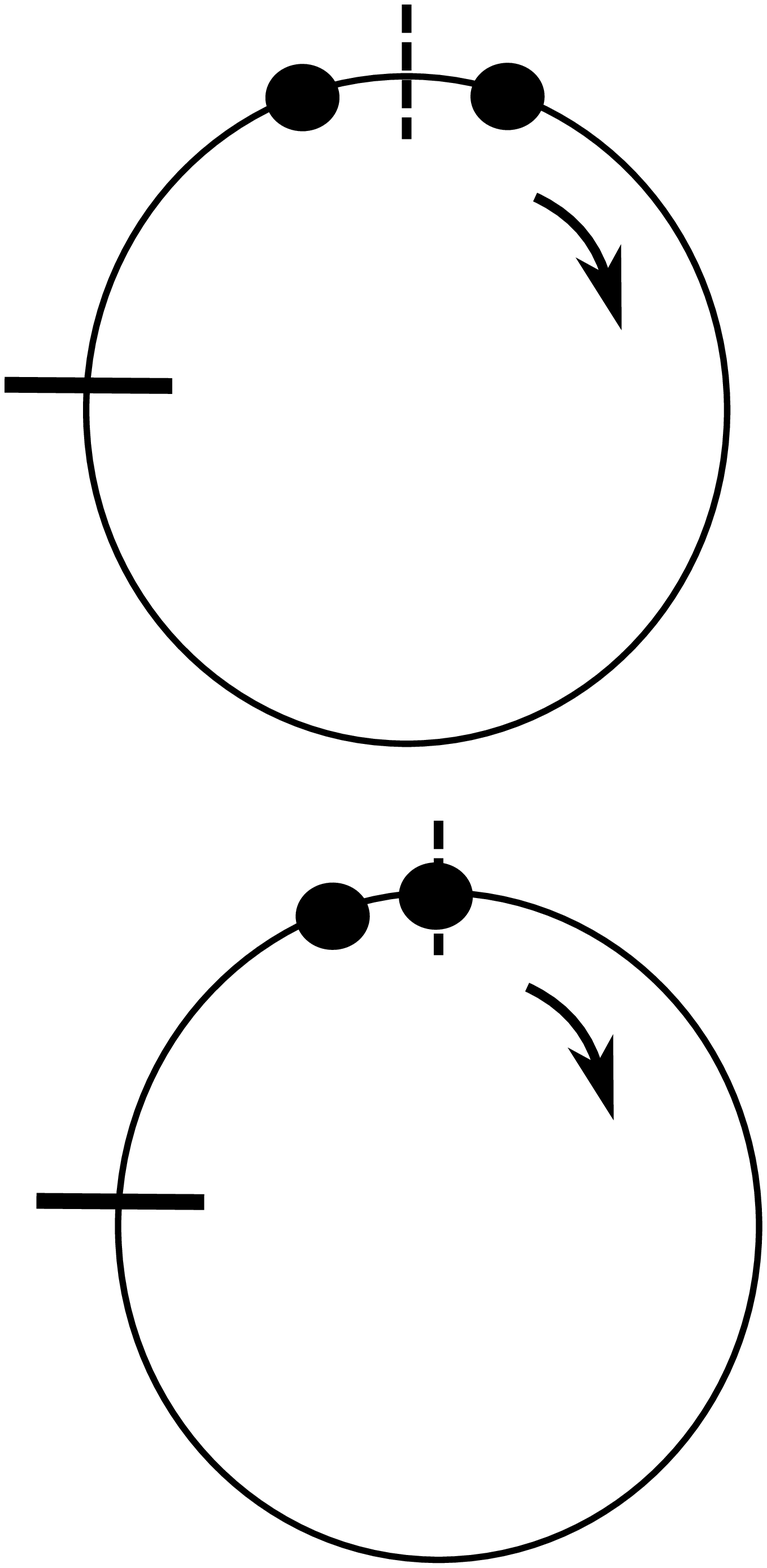}}\caption{This figure represents the
configurations we used to calculate the discrete derivative. The circle
represents the time direction of our lattice. The thicks represent monopole
and anti-monopole terms included in the action. The solid points represent the
time slices used to measure the operator and construct the derivatives. On the
left hand side we see the symmetries of the construction when interchanging monopole and antimonopole derivatives. On the right hand side differences between symmetric and backward derivatives are sketched. \label{schema}}
\end{center}
\end{figure}

To calculate the derivative of \mbox{$ \log \mathcal{O}( t )-\beta  S(t)$} numerically  in principle we could choose the  forward,  backward or symmetric prescriptions.
We are considering, however, the approximation \mref{approx} with $f(t)=\partial_{t} \left( \log
\mathcal{O}( t )-S(t)\right)$ and hence need to minimize  $f(t)$.  This requirement immediately selects  the symmetric prescription $f'(t)= (f(t+1)-f(t-1))/2.$
The reason for that is outlined 
in the right part of figure \ref{schema}. There, we sketch the scenario used in our  computations. The circle represents the temporal
direction of the lattice (the other directions are omitted for simplicity) 
and the dashed and solid thicks represent the insertion of monopole and
anti-monopole operators in the action \mref{actprim} at the two  time slices separated by
a distance $t$. The solid dots represent the time slices used to calculate
the derivative in the symmetric case and in the back-ward case. Symmetric derivatives  involve  time slices  separated by two lattice spacings.
 Nevertheless, their plaquettes  are updated with the same action and are hence more correlated than the
adjacent time slices involved in the  backward or
forward derivatives  calculations. These last  are updated with different actions since on one of them there is the monopole (or antimonopole) contribution. The symmetric prescription, hence, minimizes $\partial_{t} \left( \log
\mathcal{O}( t)-S(t)\right)$ and  is used to extract the correct value of $\rho'$. 

In order to  obtain  \mref{ron} from \mref{mu} we have a further freedom: we can derive the correlator  either with  respect to the monopole  or  to the antimonopole position, the only difference being an overall minus sign. This is
shown on the left part of figure \ref{schema}. Derivatives on the monopole field amount to minus
derivatives on the anti-monopole field.

\section{Numerical results}

\begin{figure}\begin{center}
\resizebox{7.5cm}{!}{
\epsfig{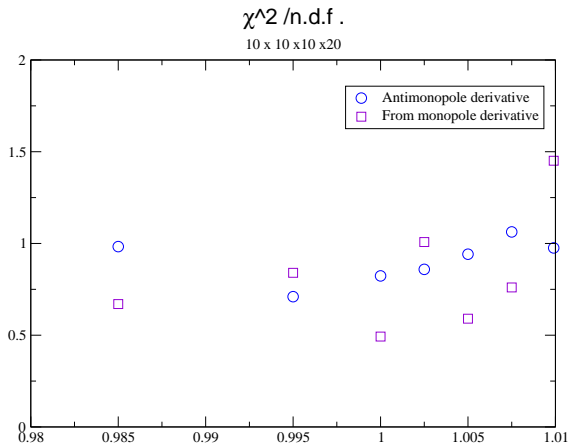}}
\caption{This figure is a plot of the $\chi^2/n.d.f.$ for the fits we performed with expression \mref{derlog} on the lattice $10^3\times 20$. For each $\beta$ we have at least one channel with $\chi^2/n.d.f.\ll 1$. This confirms the validity of \mref{derlog} to describe the correlator decay. \label{chiquadfig}}\end{center}
\end{figure}

\begin{figure}
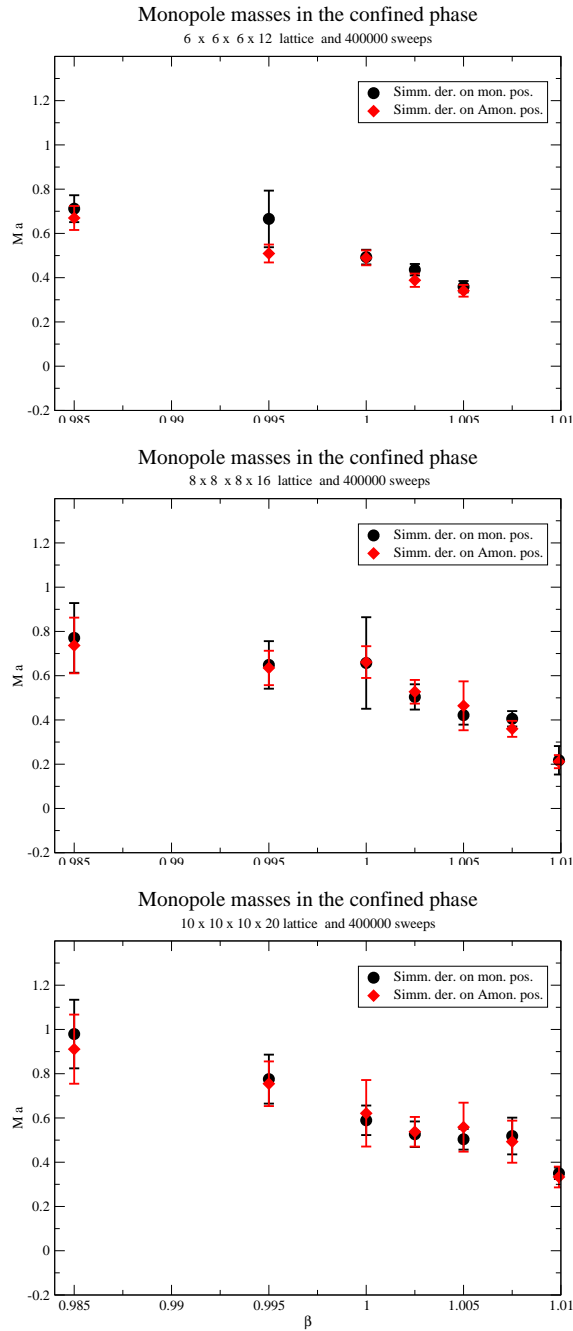
\begin{center}
\resizebox{7.5cm}{!}{
\epsfig{file=immagini/massa_log_6n.eps}}

\resizebox{7.5cm}{!}{
\epsfig{file=immagini/massa_log_8n.eps}}

\resizebox{7.5cm}{!}{
\epsfig{file=immagini/massa_log_10n.eps}}

\caption{These figures collect masses calculated with the two  possible strategies on
the three lattice sizes we considered in a range of $\beta$ in the confined phase.
We can  derive the correlation function either with respect to the monopole position
or to the anti-monopole position. Masses extracted with any of these prescriptions are compatible within error bars. \label{masses}}\end{center}
\end{figure}

We  performed three  runs of simulations on three different lattices. The
smallest is a $6^3 \times 12$ lattice the medium is a $8^3 \times 16$ and the
largest $10^3 \times 20.$ We collected over 400000 sweeps of statistics for all the  lattices . We used the
standard 1 heat-bath and 3 over-relaxation algorithm. In order to extract the value of the masses from the numerical data,  we have to include the effect of  periodic boundary conditions (p.b.c.). Once we define $f(x)=A M^{1/2}x^{-3/2}e^{-Mx}$, the p.b.c. modify \mref{dec} producing:
\begin{equation}
\langle \mu ( \vec{x}, t_1 ) \bar{\mu} ( \vec{x}, t_2 ) \rangle \sim \mu^2 +f(r)+f(r_t)+3\left(f(r_s)+f(r_{st})+f(r_{ss})+f(r_{sst})\right).\label{full}
\end{equation}
where $r_{t,} r_s, r_{s s}$ take into account the different path we can chose
to go from $( \vec{x}, t_1 )$ to $( \vec{x}, t_2 ) .$ We can in fact chose the
shortest path ($r = t = |t_2 - t_1 |$)  or the path that winds once in the
temporal direction ($r_{_t} = L - t$); we can also wind once around the
spatial direction ($r_s = \sqrt{t^2 + L^2_s}$, we have three different
choices), twice around the spatial direction ($r_{ss} = \sqrt{t^2 + ( 2 L_s
)^2}$ we have three possible choices)  once around the spatial and  the time
direction ($r_{st} = \sqrt{( L - t )^2 + L_s^2}$ we have three possible
choices) and  twice around the spatial direction and one around the time
direction  ($r_{sst} = \sqrt{( L - t )^2 + ( 2 L_s )^2}$ we have three possible
choices). Taking into account all these possibilities  we extracted the masses depicted in figure
\ref{masses} in a range of $\beta$  which ensured the system is in  the confined phase.
The first thing we notice form figure \ref{chiquadfig} is that, for all the values of $\beta$ considered, the $\chi^2 / n.d.f.$  obtained fitting the numerical data with expression \mref{derlog} is lower than one. The only exceptions are, indeed, points very close to the transition (which is  first order with very long correlation length) and are  surely due to metastabilities caused by our updating algorithm.
This means that the leading contribution to the decay is 
correctly described by a single particle excitation.
As explained in the previous section,  we use the symmetric prescription for the derivative in expression \mref{numexpr}  and 
derive both  with respect to the monopole and anti-monopole position . From the figure \ref{masses} it is clear that 
the masses  extracted in both cases   are compatible within error-bars.

\begin{figure}\begin{center}
\resizebox{8cm}{!}{\epsfig{file=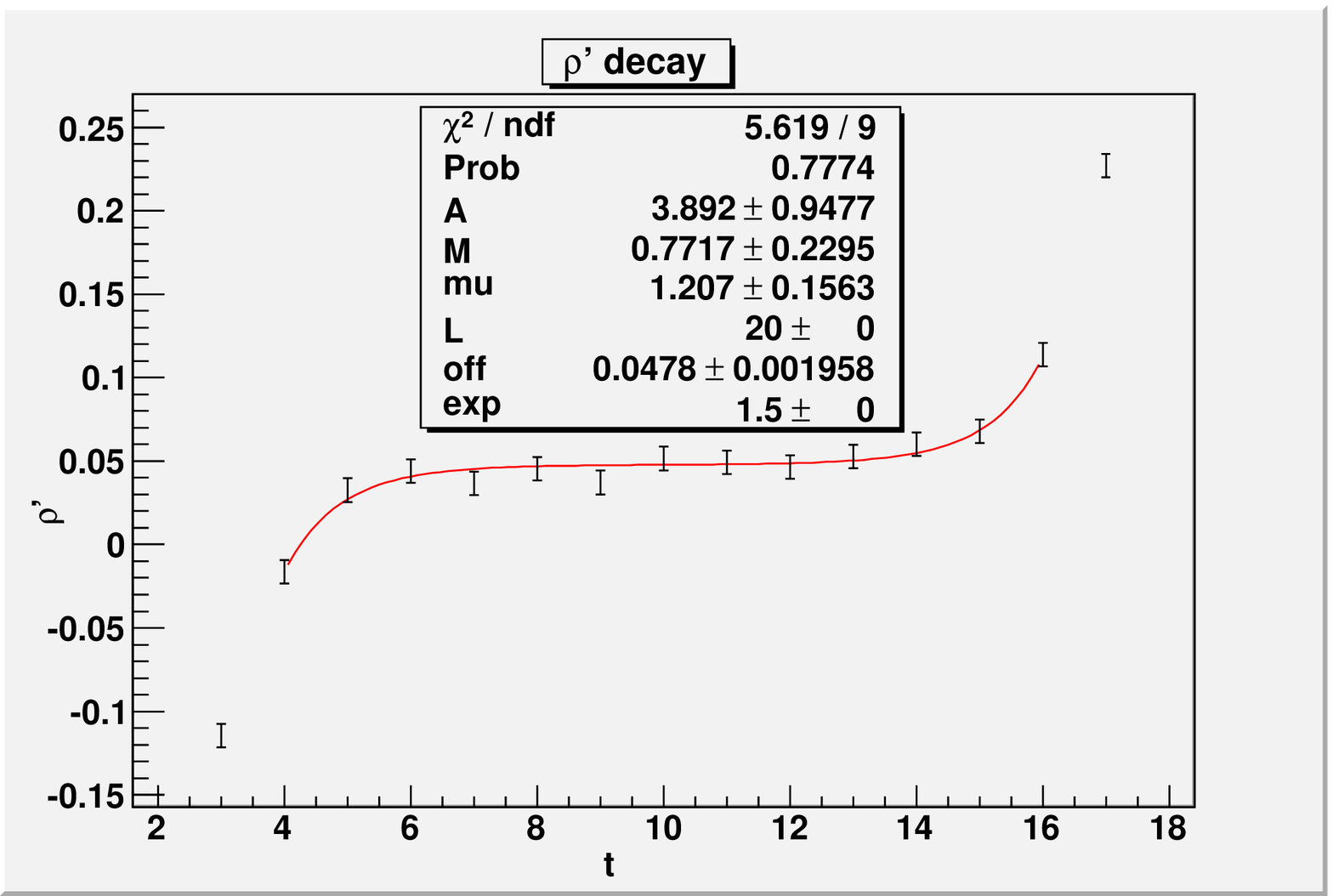}}

\resizebox{8cm}{!}{\epsfig{file=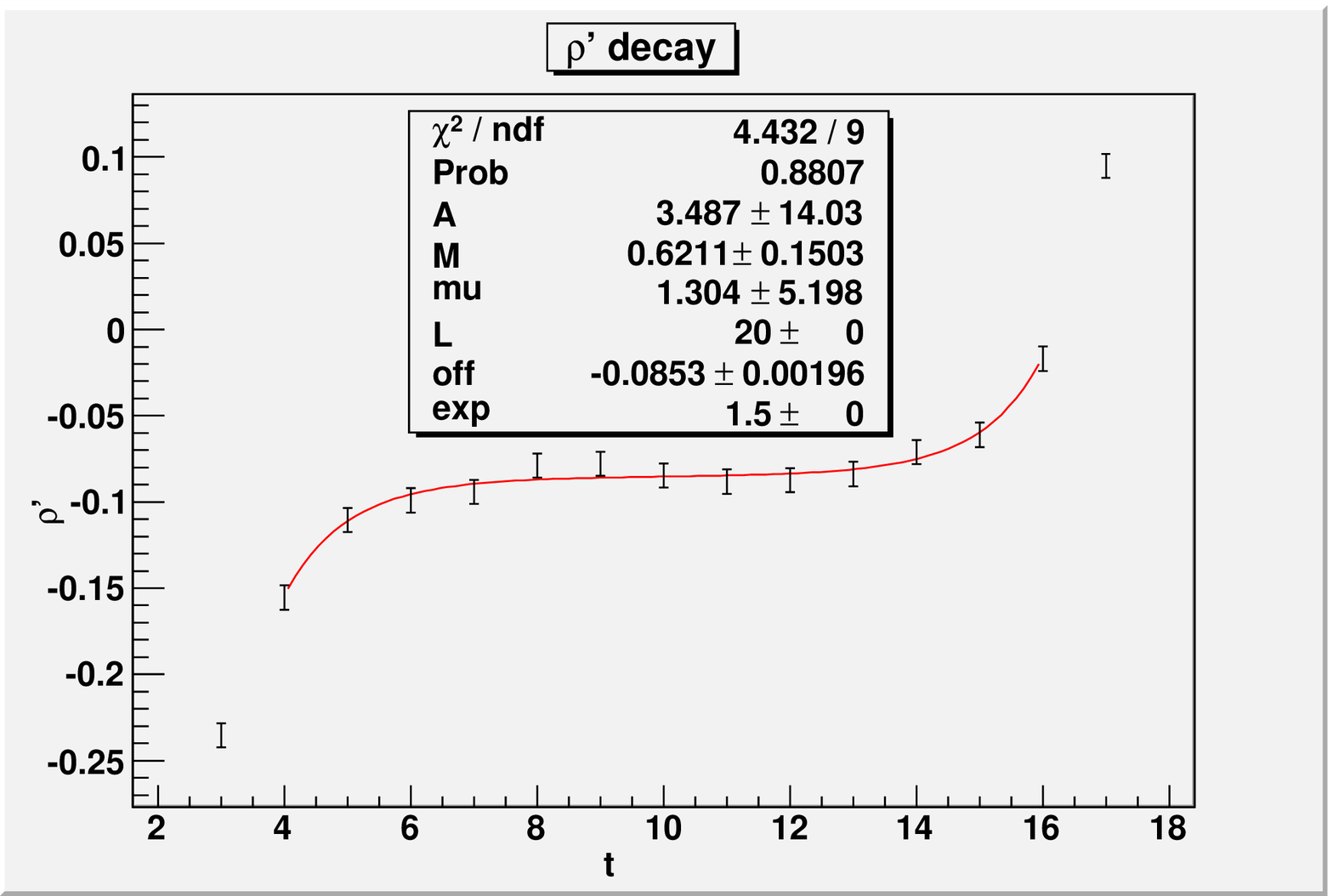}}\caption{This
figure represents the correlation function decay for a $10^3 \times 20$ lattice
at $\beta = 1.$ The points on the upper plot are obtained deriving with respect to the monopole field, while the point on the lower part deriving with respect to the anti-monopole field .\label{esfit}}
\end{center}
\end{figure}

In  figure \ref{esfit} we
show  typical data-points we obtain for the derivative of the correlation
function \mref{numexpr} from the simulations performed. The line represents the best-fit
curve. The upper  plot contains points obtained deriving with respect to the monopole 
position whether the lower plot contains  points obtained deriving with respect to the antimonopole position.
In both cases we where forced to add a 
constant term  to the expression \mref{derlog} to correct from systematic errors
induced by discretization and by expanding the exponential of the derivative in \mref{approx}. The best-fit curves
are obtained  using a subset of distances  (from 4 to 16) to avoid contamination from higher states in the same channel and safely consider only the  single particle decay. 

Comparing our results with the ones in  \cite{DiGiacomo:1997sm} one notices a clear improvement of the signal to noise ratio.
The difference between the values of the masses we extract and  the ones obtained in  \cite{DiGiacomo:1997sm} is probably due to  
the high  noise that prevented  the authors of  \cite{DiGiacomo:1997sm} from using  the full expression \mref{full} in fits.
The authors, forced to use a zero momentum approximation of the decay \mref{dec},  introduced a systematic error that shifted the masses from their actual value.

\begin{figure}\begin{center}
\resizebox{8cm}{!}{\epsfig{file=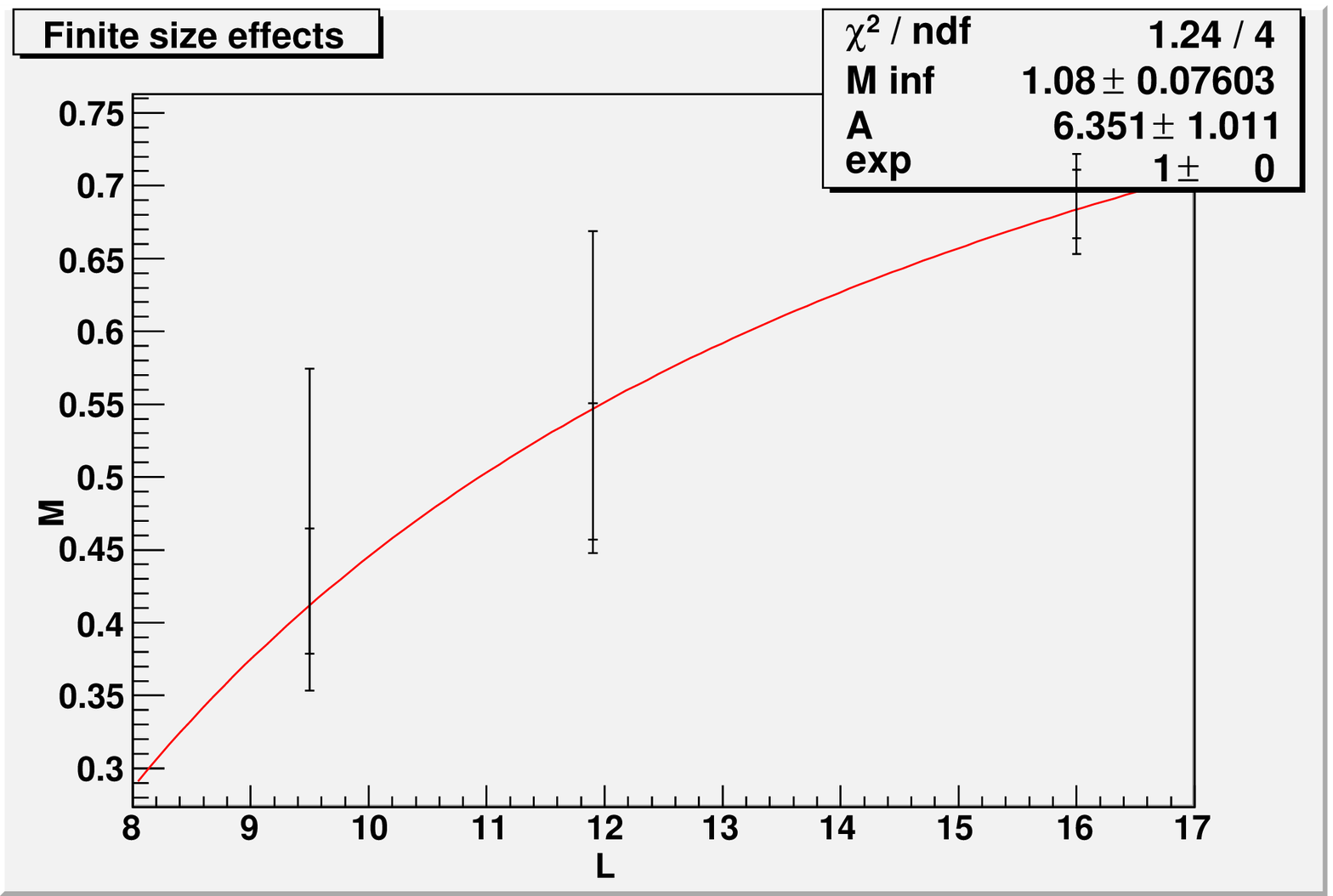}}

\resizebox{8cm}{!}{\epsfig{file=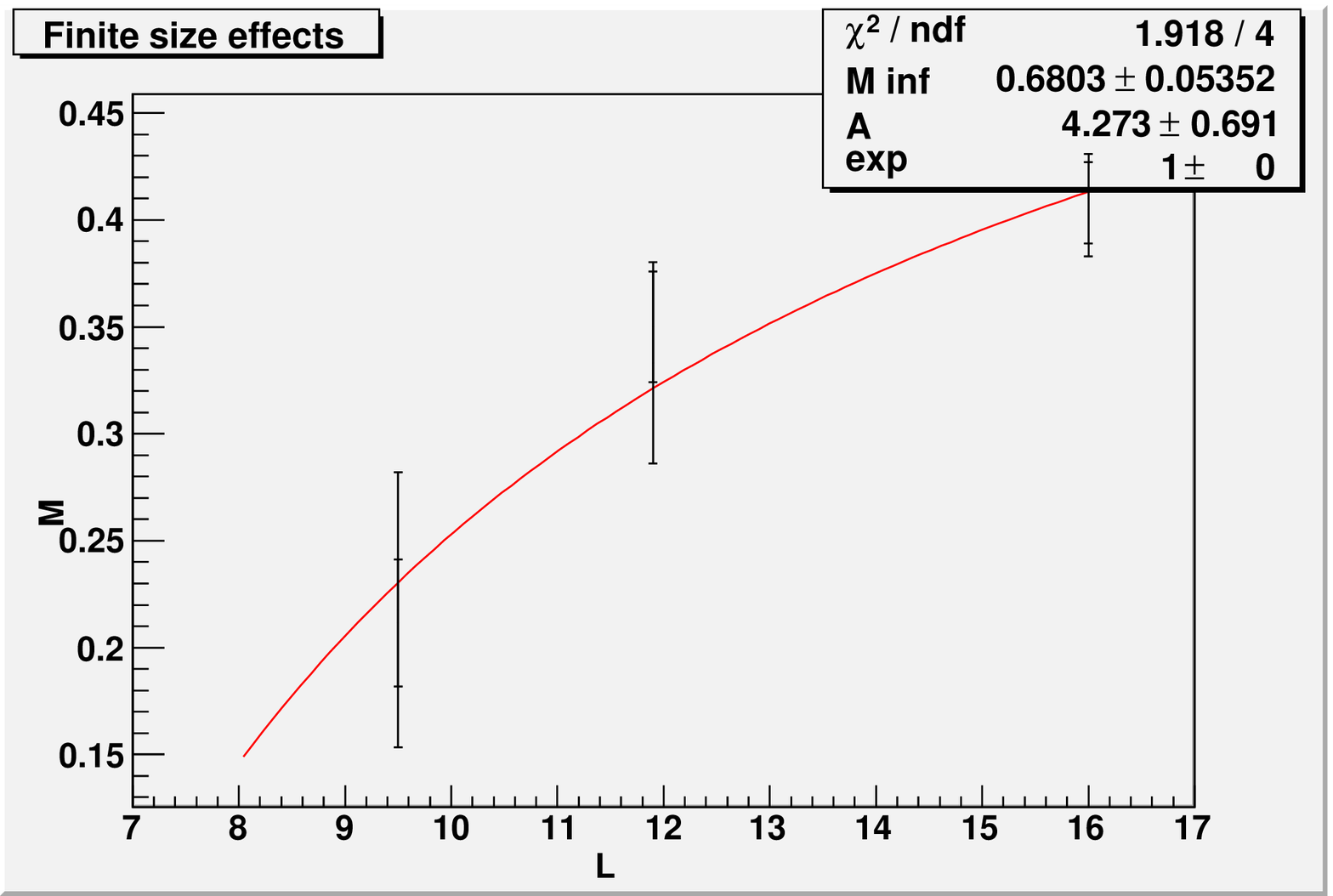}}\caption{These plots show the test of the assumption that the discrepancies between our results and the one contained in  {\cite{Majumdar:2003xm}} for the $0^{+ +}$ mass are  finite size effects. 
We plot the mass against the linear size of the lattice for two different choices of $\beta$ close to the transition. The last point in both plots (the one at $L=16$) is the $0^{+ +}$ mass taken from  \cite{Majumdar:2003xm}. The upper plot is at $\beta=1.005$.
The lower plot is at $\beta=1.0099$ .We considered  only the two bigger lattices as for them the scaling should be dominated by $1/L$ effects.
We considered jointly   results extracted from both monopole and  antimonopole channels.
The  values of the $\chi^2$ we obtain from the fit confirm the identification of the state we are studying with the  $0^{+ +}$ gaugeball. \label{fs}}
\end{center}
\end{figure}

We can also safely identify the particle responsible for the decay. The masses we get from the fit with the complete expression
\mref{full} are indeed  fully compatible with the  known value for the gaugeball
$0^{+ +}$ mass (see i.e.  {\cite{Majumdar:2003xm}} for a recent high precision
study).
To show this we took into account finite size effects in the two plots \ref{fs}.

\begin{figure}\begin{center}
\resizebox{8cm}{!}{\epsfig{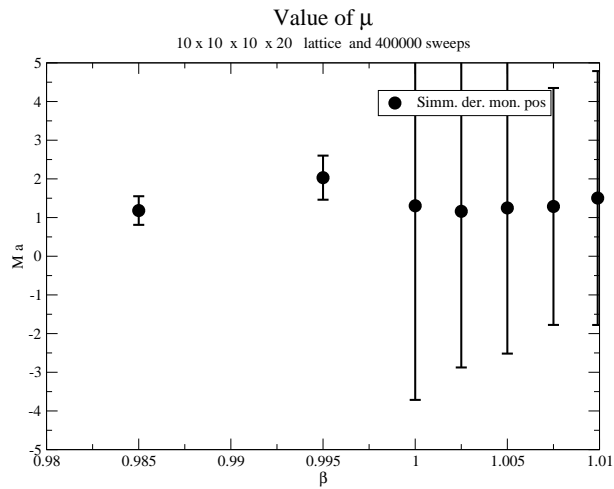}}\caption{V.e.v. of $\mu$ extracted from our definition of $\rho'$. This plot shows that this operator is not adequate to measure the order parameter for the Coulomb-Confined transition.\label{mufig}}
\end{center}
\end{figure}

In the first we considered  the results  we obtained  for the masses (from both monopole and anti-monopole channels) at $\beta=1.005$. We   plotted them  versus the lattice size (calculated as  $L=\left(L_s^3 \times L_t\right)^{1/4}$). The last point at $L=16$ is taken from the results of the $0^{+ +}$ mass contained in  {\cite{Majumdar:2003xm}}. Under the assumption that the discrepancy between our results and the one {\cite{Majumdar:2003xm}} is due to finite size effects we made a two parameters fit with the expression $M(L)=M_{\infty}-A L^{-1}$.
In the second plot there is the same study  at $\beta=1.0099$.
The $\chi^2$ values safely confirm the identification of the particle responsible for the decay \mref{dec}  with  the $0^{+ +}$ gaugeball.  
Furthermore the fact that the thermodynamic limit is approached from below is the expected behaviour in the confined phase (see i.e.  \cite{Montvay:1994cy}).
These results confirm  the validity of a theorem about the structure of the Hilbert
space in the confined phase of compact lattice $U ( 1 )$ in four dimensions proved in
 {\cite{Frohlich:1986sz}}: there are  no super-selected sectors labelled
by the magnetic charge.
The last point to stress is that our method prevents a precise measurement of the v.e.v.  $\mu$   (as it is designed to get rid of it as far as possible). This  is shown in figure \ref{mufig} and implies that, if  interested in studying the order parameter for the Coulomb Confined transition, one should better use the standard  $\rho$ definition 
\cite{DelDebbio:1994sx}.

\section{Conclusions}

In this work we considered the spectrum extracted from magnetically charged
operators. We introduced a new powerful technique that reduces, with respect to existing methods,  one order of magnitude
 the amount of data necessary to extract a good signal for  correlations
of such operators .

 The improvement of the technique relies on the fact that it allows  to study the connected
part of the correlation functions. As a first application we  study the
Hilbert space of confined compact four dimensional $U ( 1 )$ theory. We
consider monopole anti-monopole correlation functions on three lattice sizes
($6^3 \times 12$, $8^3 \times 16$, $10^3 \times 20$). We   check that the 
large time correlation functions decay is driven by a single particle. We also identify this
particle by studying  its mass. The value for the mass we extract is compatible with the
one found in literature for the gaugeball $0^{+ +}$. This is in complete
agreement with a theorem proved in  {\cite{Frohlich:1986sz}} stating that
the Hilbert space of compact lattice $U ( 1 )$ in four dimensions does not
contains any magnetically super-selected sectors. Completing our results
with the ones obtained by other groups with different techniques
{\cite{Majumdar:2003xm,Panero:2005iu,Koma:2003gi}} one
can have a full picture of the Hilbert space of compact $U ( 1 )$ lattice
gauge theory in four dimensions as the space spanned by gaugeball states. We are
working on similar studies for the case of non Abelian lattice gauge theories.
We are also working on  the implementation 
of an new algorithm to measure directly the order parameter $\vec{\mu}(\vec x,t)$ and its finite size scaling  based on ideas similar to the ones presented in  this work.

\section{Acknowledgements}

The work was partially financed by a graduate scholarship of the Spanish M.E.C. I
would like to acknowledge D. Espriu, A. Di Giacomo, F. Gliozzi, G. Paffuti and
C. Pica for the precious discussions and suggestions on the topics covered by
this work. I also want to acknowledge M. D'Elia for correcting an error in a preliminary version
of this work and the Physics Department at the Pisa University for its hospitality during the earlier stage
of this work.


\begin{thebibliography}{26}
\expandafter\ifx\csname natexlab\endcsname\relax\def\natexlab#1{#1}\fi
\expandafter\ifx\csname bibnamefont\endcsname\relax
  \def\bibnamefont#1{#1}\fi
\expandafter\ifx\csname bibfnamefont\endcsname\relax
  \def\bibfnamefont#1{#1}\fi
\expandafter\ifx\csname citenamefont\endcsname\relax
  \def\citenamefont#1{#1}\fi
\expandafter\ifx\csname url\endcsname\relax
  \def\url#1{\texttt{#1}}\fi
\expandafter\ifx\csname urlprefix\endcsname\relax\def\urlprefix{URL }\fi
\providecommand{\bibinfo}[2]{#2}
\providecommand{\eprint}[2][]{\url{#2}}

\bibitem[{\citenamefont{Wilson}(1974)}]{Wilson:1974sk}
\bibinfo{author}{\bibfnamefont{K.~G.} \bibnamefont{Wilson}},
  \bibinfo{journal}{Phys. Rev.} \textbf{\bibinfo{volume}{D10}},
  \bibinfo{pages}{2445} (\bibinfo{year}{1974}).

\bibitem[{\citenamefont{Balian et~al.}(1974)\citenamefont{Balian, Drouffe, and
  Itzykson}}]{Balian:1974ts}
\bibinfo{author}{\bibfnamefont{R.}~\bibnamefont{Balian}},
  \bibinfo{author}{\bibfnamefont{J.~M.} \bibnamefont{Drouffe}},
  \bibnamefont{and} \bibinfo{author}{\bibfnamefont{C.}~\bibnamefont{Itzykson}},
  \bibinfo{journal}{Phys. Rev.} \textbf{\bibinfo{volume}{D10}},
  \bibinfo{pages}{3376} (\bibinfo{year}{1974}).

\bibitem[{\citenamefont{Balian et~al.}(1975)\citenamefont{Balian, Drouffe, and
  Itzykson}}]{Balian:1974ir}
\bibinfo{author}{\bibfnamefont{R.}~\bibnamefont{Balian}},
  \bibinfo{author}{\bibfnamefont{J.~M.} \bibnamefont{Drouffe}},
  \bibnamefont{and} \bibinfo{author}{\bibfnamefont{C.}~\bibnamefont{Itzykson}},
  \bibinfo{journal}{Phys. Rev.} \textbf{\bibinfo{volume}{D11}},
  \bibinfo{pages}{2098} (\bibinfo{year}{1975}).

\bibitem[{\citenamefont{Creutz et~al.}(1979)\citenamefont{Creutz, Jacobs, and
  Rebbi}}]{Creutz:1979zg}
\bibinfo{author}{\bibfnamefont{M.}~\bibnamefont{Creutz}},
  \bibinfo{author}{\bibfnamefont{L.}~\bibnamefont{Jacobs}}, \bibnamefont{and}
  \bibinfo{author}{\bibfnamefont{C.}~\bibnamefont{Rebbi}},
  \bibinfo{journal}{Phys. Rev.} \textbf{\bibinfo{volume}{D20}},
  \bibinfo{pages}{1915} (\bibinfo{year}{1979}).

\bibitem[{\citenamefont{Lautrup and Nauenberg}(1980)}]{Lautrup:1980xr}
\bibinfo{author}{\bibfnamefont{B.}~\bibnamefont{Lautrup}} \bibnamefont{and}
  \bibinfo{author}{\bibfnamefont{M.}~\bibnamefont{Nauenberg}},
  \bibinfo{journal}{Phys. Lett.} \textbf{\bibinfo{volume}{B95}},
  \bibinfo{pages}{63} (\bibinfo{year}{1980}).

\bibitem[{\citenamefont{Evertz et~al.}(1985)\citenamefont{Evertz, Jersak,
  Neuhaus, and Zerwas}}]{Evertz:1984pn}
\bibinfo{author}{\bibfnamefont{H.~G.} \bibnamefont{Evertz}},
  \bibinfo{author}{\bibfnamefont{J.}~\bibnamefont{Jersak}},
  \bibinfo{author}{\bibfnamefont{T.}~\bibnamefont{Neuhaus}}, \bibnamefont{and}
  \bibinfo{author}{\bibfnamefont{P.~M.} \bibnamefont{Zerwas}},
  \bibinfo{journal}{Nucl. Phys.} \textbf{\bibinfo{volume}{B251}},
  \bibinfo{pages}{279} (\bibinfo{year}{1985}).

\bibitem[{\citenamefont{Arnold et~al.}(2001)\citenamefont{Arnold, Lippert,
  Schilling, and Neuhaus}}]{Arnold:2000hf}
\bibinfo{author}{\bibfnamefont{G.}~\bibnamefont{Arnold}},
  \bibinfo{author}{\bibfnamefont{T.}~\bibnamefont{Lippert}},
  \bibinfo{author}{\bibfnamefont{K.}~\bibnamefont{Schilling}},
  \bibnamefont{and} \bibinfo{author}{\bibfnamefont{T.}~\bibnamefont{Neuhaus}},
  \bibinfo{journal}{Nucl. Phys. Proc. Suppl.} \textbf{\bibinfo{volume}{94}},
  \bibinfo{pages}{651} (\bibinfo{year}{2001}), \eprint{hep-lat/0011058}.

\bibitem[{\citenamefont{Vettorazzo and de~Forcrand}(2004)}]{Vettorazzo:2003fg}
\bibinfo{author}{\bibfnamefont{M.}~\bibnamefont{Vettorazzo}} \bibnamefont{and}
  \bibinfo{author}{\bibfnamefont{P.}~\bibnamefont{de~Forcrand}},
  \bibinfo{journal}{Nucl. Phys.} \textbf{\bibinfo{volume}{B686}},
  \bibinfo{pages}{85} (\bibinfo{year}{2004}), \eprint{hep-lat/0311006}.

\bibitem[{\citenamefont{Polyakov}(1975)}]{Polyakov:1975rs}
\bibinfo{author}{\bibfnamefont{A.~M.} \bibnamefont{Polyakov}},
  \bibinfo{journal}{Phys. Lett.} \textbf{\bibinfo{volume}{B59}},
  \bibinfo{pages}{82} (\bibinfo{year}{1975}).

\bibitem[{\citenamefont{Mandelstam}(1975)}]{Mandelstam:1974vf}
\bibinfo{author}{\bibfnamefont{S.}~\bibnamefont{Mandelstam}},
  \bibinfo{journal}{Phys. Lett.} \textbf{\bibinfo{volume}{B53}},
  \bibinfo{pages}{476} (\bibinfo{year}{1975}).

\bibitem[{\citenamefont{'t~Hooft}(1975)}]{'tHooft:1975pr}
\bibinfo{author}{\bibfnamefont{G.}~\bibnamefont{'t~Hooft}}
  (\bibinfo{year}{1975}), \bibinfo{note}{lectures given at Int. School of
  Subnuclear Physics, 'Ettore Majorana', Erice, Sicily, Jul 11-31}.

\bibitem[{\citenamefont{DeGrand and Toussaint}(1980)}]{DeGrand:1980eq}
\bibinfo{author}{\bibfnamefont{T.~A.} \bibnamefont{DeGrand}} \bibnamefont{and}
  \bibinfo{author}{\bibfnamefont{D.}~\bibnamefont{Toussaint}},
  \bibinfo{journal}{Phys. Rev.} \textbf{\bibinfo{volume}{D22}},
  \bibinfo{pages}{2478} (\bibinfo{year}{1980}).

\bibitem[{\citenamefont{Frohlich and Marchetti}(1987)}]{Frohlich:1987er}
\bibinfo{author}{\bibfnamefont{J.}~\bibnamefont{Frohlich}} \bibnamefont{and}
  \bibinfo{author}{\bibfnamefont{P.~A.} \bibnamefont{Marchetti}},
  \bibinfo{journal}{Commun. Math. Phys.} \textbf{\bibinfo{volume}{112}},
  \bibinfo{pages}{343} (\bibinfo{year}{1987}).

\bibitem[{\citenamefont{Polley and Wiese}(1991)}]{Polley:1990tf}
\bibinfo{author}{\bibfnamefont{L.}~\bibnamefont{Polley}} \bibnamefont{and}
  \bibinfo{author}{\bibfnamefont{U.~J.} \bibnamefont{Wiese}},
  \bibinfo{journal}{Nucl. Phys.} \textbf{\bibinfo{volume}{B356}},
  \bibinfo{pages}{629} (\bibinfo{year}{1991}).

\bibitem[{\citenamefont{Del~Debbio et~al.}(1995)\citenamefont{Del~Debbio,
  Di~Giacomo, and Paffuti}}]{DelDebbio:1994sx}
\bibinfo{author}{\bibfnamefont{L.}~\bibnamefont{Del~Debbio}},
  \bibinfo{author}{\bibfnamefont{A.}~\bibnamefont{Di~Giacomo}},
  \bibnamefont{and} \bibinfo{author}{\bibfnamefont{G.}~\bibnamefont{Paffuti}},
  \bibinfo{journal}{Phys. Lett.} \textbf{\bibinfo{volume}{B349}},
  \bibinfo{pages}{513} (\bibinfo{year}{1995}), \eprint{hep-lat/9403013}.

\bibitem[{\citenamefont{Di~Giacomo and Paffuti}(1997)}]{DiGiacomo:1997sm}
\bibinfo{author}{\bibfnamefont{A.}~\bibnamefont{Di~Giacomo}} \bibnamefont{and}
  \bibinfo{author}{\bibfnamefont{G.}~\bibnamefont{Paffuti}},
  \bibinfo{journal}{Phys. Rev.} \textbf{\bibinfo{volume}{D56}},
  \bibinfo{pages}{6816} (\bibinfo{year}{1997}), \eprint{hep-lat/9707003}.

\bibitem[{\citenamefont{Jersak et~al.}(1999)\citenamefont{Jersak, Neuhaus, and
  Pfeiffer}}]{Jersak:1999nv}
\bibinfo{author}{\bibfnamefont{J.}~\bibnamefont{Jersak}},
  \bibinfo{author}{\bibfnamefont{T.}~\bibnamefont{Neuhaus}}, \bibnamefont{and}
  \bibinfo{author}{\bibfnamefont{H.}~\bibnamefont{Pfeiffer}},
  \bibinfo{journal}{Phys. Rev.} \textbf{\bibinfo{volume}{D60}},
  \bibinfo{pages}{054502} (\bibinfo{year}{1999}), \eprint{hep-lat/9903034}.

\bibitem[{\citenamefont{Seiberg and Witten}(1994)}]{Seiberg:1994rs}
\bibinfo{author}{\bibfnamefont{N.}~\bibnamefont{Seiberg}} \bibnamefont{and}
  \bibinfo{author}{\bibfnamefont{E.}~\bibnamefont{Witten}},
  \bibinfo{journal}{Nucl. Phys.} \textbf{\bibinfo{volume}{B426}},
  \bibinfo{pages}{19} (\bibinfo{year}{1994}),
  \eprint[http://arXiv.org/abs]{hep-th/9407087}.

\bibitem[{\citenamefont{Espriu and Tagliacozzo}(2003)}]{Espriu:2003sa}
\bibinfo{author}{\bibfnamefont{D.}~\bibnamefont{Espriu}} \bibnamefont{and}
  \bibinfo{author}{\bibfnamefont{L.}~\bibnamefont{Tagliacozzo}},
  \bibinfo{journal}{Phys. Lett.} \textbf{\bibinfo{volume}{B557}},
  \bibinfo{pages}{125} (\bibinfo{year}{2003}), \eprint{hep-th/0301086}.

\bibitem[{\citenamefont{Espriu and Tagliacozzo}(2004)}]{Espriu:2004ib}
\bibinfo{author}{\bibfnamefont{D.}~\bibnamefont{Espriu}} \bibnamefont{and}
  \bibinfo{author}{\bibfnamefont{L.}~\bibnamefont{Tagliacozzo}},
  \bibinfo{journal}{Phys. Lett.} \textbf{\bibinfo{volume}{B602}},
  \bibinfo{pages}{137} (\bibinfo{year}{2004}), \eprint{hep-th/0405015}.

\bibitem[{\citenamefont{Frohlich and Marchetti}(1986)}]{Frohlich:1986sz}
\bibinfo{author}{\bibfnamefont{J.}~\bibnamefont{Frohlich}} \bibnamefont{and}
  \bibinfo{author}{\bibfnamefont{P.~A.} \bibnamefont{Marchetti}},
  \bibinfo{journal}{Europhys. Lett.} \textbf{\bibinfo{volume}{2}},
  \bibinfo{pages}{933} (\bibinfo{year}{1986}).

\bibitem[{\citenamefont{Majumdar et~al.}(2004)\citenamefont{Majumdar, Koma, and
  Koma}}]{Majumdar:2003xm}
\bibinfo{author}{\bibfnamefont{P.}~\bibnamefont{Majumdar}},
  \bibinfo{author}{\bibfnamefont{Y.}~\bibnamefont{Koma}}, \bibnamefont{and}
  \bibinfo{author}{\bibfnamefont{M.}~\bibnamefont{Koma}},
  \bibinfo{journal}{Nucl. Phys.} \textbf{\bibinfo{volume}{B677}},
  \bibinfo{pages}{273} (\bibinfo{year}{2004}), \eprint{hep-lat/0309003}.

\bibitem[{\citenamefont{Cox et~al.}(1998)}]{Cox:1997nq}
\bibinfo{author}{\bibfnamefont{J.}~\bibnamefont{Cox}} \bibnamefont{et~al.},
  \bibinfo{journal}{Nucl. Phys. Proc. Suppl.} \textbf{\bibinfo{volume}{63}},
  \bibinfo{pages}{691} (\bibinfo{year}{1998}), \eprint{hep-lat/9709054}.

\bibitem[{\citenamefont{Panero}(2005)}]{Panero:2005iu}
\bibinfo{author}{\bibfnamefont{M.}~\bibnamefont{Panero}},
  \bibinfo{journal}{JHEP} \textbf{\bibinfo{volume}{05}}, \bibinfo{pages}{066}
  (\bibinfo{year}{2005}), \eprint{hep-lat/0503024}.

\bibitem[{\citenamefont{Koma et~al.}(2004)\citenamefont{Koma, Koma, and
  Majumdar}}]{Koma:2003gi}
\bibinfo{author}{\bibfnamefont{Y.}~\bibnamefont{Koma}},
  \bibinfo{author}{\bibfnamefont{M.}~\bibnamefont{Koma}}, \bibnamefont{and}
  \bibinfo{author}{\bibfnamefont{P.}~\bibnamefont{Majumdar}},
  \bibinfo{journal}{Nucl. Phys.} \textbf{\bibinfo{volume}{B692}},
  \bibinfo{pages}{209} (\bibinfo{year}{2004}), \eprint{hep-lat/0311016}.

\bibitem[{\citenamefont{Montvay and Munster}(1994)}]{Montvay:1994cy}
\bibinfo{author}{\bibfnamefont{I.}~\bibnamefont{Montvay}} \bibnamefont{and}
  \bibinfo{author}{\bibfnamefont{G.}~\bibnamefont{Munster}}
  (\bibinfo{year}{1994}), \bibinfo{note}{cambridge, UK: Univ. Pr. 491 p.
  (Cambridge monographs on mathematical physics)}.

\end{thebibliography}
\end{document}